# Subband engineering for p-type silicon ultra-thin layers for increased carrier velocities: An atomistic analysis

Neophytos Neophytou, Gerhard Klimeck* and Hans Kosina

Institute for Microelectronics, TU Wien, Gußhausstraße 27-29/E360, A-1040 Wien, Austria

e-mail: {neophytou|kosina}@iue.tuwien.ac.at

[*]Network for Computational Nanotechnology, Purdue University, IN, 47907, USA.

## Abstract

Ultra-thin-body (UTB) channel materials of a few nanometers in thickness are currently considered as candidates for future electronic, thermoelectric, and optoelectronic applications. Among the features that they possess, which make them attractive for such applications, their confinement length scale, transport direction, and confining surface orientation serve as degrees of freedom for engineering their electronic properties. This work presents a comprehensive study of hole velocities in p-type UTB films of widths from 15nm down to 3nm. Various transport and surface orientations are considered. The atomistic $sp^3d^5s^*$-spin-orbit-coupled tight-binding model is used for the electronic structure, and a semiclassical ballistic model for the carrier velocity calculation. We find that the carrier velocity is a strong function of orientation and layer thickness. The (110) and (112) surfaces provide the highest hole velocities, whereas the (100) surfaces the lowest velocities, almost 30% lower than the best performers. Additionally, up to 35% velocity enhancements can be achieved as the thickness of the (110) or (112) surfaces is scaled down to 3nm. This originates from strong increase in the curvature of the p-type UTB film subbands with confinement, unlike the case of n-type UTB channels. The velocity behavior directly translates to ballistic on-current trends, and correlates with trends in experimental mobility measurements.

# I. Introduction

Ultra-thin-body (UTB) channel devices have recently attracted significant attention as candidates for a variety of applications. For high performance electronic applications, silicon UTB channels offer the possibility of enhanced electrostatic control and are already being considered for future CMOS devices [1]. Thin layer superlattices and other low dimensional channels have also attracted attention for possible thermoelectric applications with enhanced performance because of their potentially improved power factors [2], and lower thermal conductivity [3, 4, 5, 6]. Lower dimensionality also seems to attract attention for applications in optoelectronics [7, 8] and biosensing [9, 10].

At the nanoscale, enhanced electron/hole confinement can alter the dispersion and electronic properties of a channel material. The confinement length scale, transport and surface orientations, could serve as additional degrees of freedom in engineering device properties. Regarding different orientations, mobility measurements and calculations on some specific silicon-on-insulator (SOI), UTB geometries, and MOSFET devices indicated orientation dependence for both, n-type [11, 12], but even more evidently for p-type channels [12, 13, 14]. Regarding the influence of confinement, we previously showed that the average carrier velocity in silicon nanowires (NWs) can in certain cases strongly vary as the diameter scales below 12nm [15]. This is again, especially evident in the case of p-type NW channels [15].

In Ref. [15] we performed a comprehensive analysis of the geometry influence on the carrier velocity of n-type and p-type NWs and provided explanations based on their electronic structure. NWs, however, are subject to 2D confinement and the influence of all confining surfaces is intermixed. The ability to identify the influence of each confining surface independently, is more important for UTB channels with a single confinement, and can act as a design parameter. In Ref. [11] we examined the geometry dependence of electron velocity for some cases of n-type silicon UTB channels. For n-type channels, such effects can be understood through the transformation of the transport and confinement effective masses of the six-fold degenerate valleys upon quantization, and the relative energy shift of those valleys [11, 16, 17]. Extensive literature is available on the properties of silicon n-type nanoscale devices.

Studies on the influence of geometry on p-type UTB channels are, however, limited mostly due to the computational complexities involved in rigorously treating the valence band. In p-type channels the geometry affects the electronic properties by causing strong changes to the subbands' curvature [15, 18, 19], unlike the case of n-type channels [11, 15]. To capture that, appropriate simulation methods are needed, beyond the effective mass approximation such as tight-binding [18, 19, 20, 21] or k•p methods [13, 22]. A comprehensive study that investigates the effect of confinement and orientation on the carrier velocities of p-type UTB channels, and identifies the underlying bandstructure mechanisms responsible, has not yet been reported, and it is the subject of this work. For this, we use the atomistic $sp^3d^5s^*$-spin-orbit-coupled tight-binding model for the electronic structure [23, 24, 25], and a semiclassical ballistic model [26, 27] for transport calculations.

We present a complete study of the hole velocities in p-type Si UTB films as a function of: i) transport orientation, ii) confining surface orientation as shown in Fig. 1, and iii) layer thickness from $W$=15nm down to $W$=3nm. We find that hole velocities are highly anisotropic with respect to transport and quantization orientations, as also suggested by mobility measurements [12, 28]. Thickness scaling can in certain cases increase the carrier velocities significantly, just as in the case of p-type NWs [15]. Unlike in NWs where all-around surfaces are simultaneously confined, here the influence of each surface can be identified. Different surfaces can have different effect on the dispersion. The best performer is the (1-10)/[110] channel. Strong confinement of the (1-10) surface down to $W$=3nm can increase the hole velocities by ~35%. On the other hand, the worst performer is the (001)/[110] UTB channel. This demonstrates the importance of the confining surface over the transport orientation. The velocity trends translate to ballistic on-currents and agree well with trends in mobility measurements [12, 28]. Finally, we compare the velocities of the p-type UTB films to the corresponding n-type ones. The velocities of n-type channels have: i) weaker geometry dependence and ii) reverse magnitude ordering with respect to orientation compared to p-type channels. Our results could provide insight into understanding recent mobility measurements [12, 14], as well as design optimization directions.

## II. Approach

The UTB channels' bandstructure is calculated using a 20 orbital atomistic tight-binding spin-orbit-coupled model ($sp^3d^5s^*$-SO) [23, 24, 25]. The channel description is built on the actual diamond lattice and each atom is properly accounted in the calculation. It accurately captures the electronic structure and the respective carrier velocities, and inherently includes the effects of quantization and different orientations. The model is a compromise between fully ab-initio models and simple effective mass methods, while being computationally affordable. The $sp^3d^5s^*$-SO model, with the parametrization of Ref. [23] was extensively used in the calculation of the electronic properties of nanostructures with excellent agreement to experimental observations [29, 30, 31]. For the calculation of transport properties and carrier velocity, a semiclassical ballistic model is used [26, 27]. In this work we examine the effect of bandstructure on the carrier velocities assuming a constant electrostatic potential in the cross sections. Our results are, therefore, strictly valid for flat-band conditions, however, the basic trends can also be observed under inversion conditions. We consider here infinitely long UTB films. No relaxation is assumed for the lattices near the surface. The electronic structure of ultra scaled devices is sensitive to the cross section size and crystal orientations [15, 19, 25]. Differences in the shapes of the dispersions between UTB channels of different orientations and widths, in the number of subbands, and their spread in energy, can result in different transport properties. To investigate these effects we consider three different transport orientations [100], [110], and [111], various confining surface orientations (as shown in Fig. 1), and film thicknesses from $W$=15nm down to $W$=3nm.

## III. Results

Quantization of different directions in a UTB film can have different effects on the electronic structure and properties. In the case of n-type UTB silicon channels under arbitrary orientation and cross section quantization, the bandstructure is mostly determined by the transport and quantization effective masses of the rotated conduction band ellipsoids [16, 17]. Although mass variations are observed under extreme cross section scaling [25], the carrier velocities are mostly determined by the well defined values for the longitudinal and transverse effective masses. In the case of the valence band, the situation is different. The anisotropic nature of the heavy-hole still results in transport and confinement orientation dependent channel dispersions. For certain cases, however, width scaling brings additional curvature variations that strongly increase the carrier velocity.

Figure 2 demonstrates the effect of surface confinement on the valence band. Figures 2a and 2b show a case in which confinement causes only weak changes in the electronic structure, whereas Fig. 2c and 2d a case in which confinement has a strong effect. Figures 2a and 2b show the 2D energy surface of the highest valence band for a UTB channel with (010) surface confinement for thicknesses $W$=15nm and $W$=3nm respectively. The contour lines (from the center outwards) show energy contours at 0.02eV, 0.05eV, 0.1eV and 0.2eV below the top of the valence band. Some anisotropic behavior is observed in the energy contours of each figure. Differences in the contours between the two figures are also observed. As we show further on, these differences are not strong enough to significantly influence the electronic characteristics of (010) confined UTB channels, either in terms of anisotropic transport, or film thickness.

Figures 2c and 2d, show the energy surface contours of the UTB film with (1-10) surface confinement with thicknesses $W$=15nm and $W$=3nm, respectively. In this case three observations can be made regarding strong contour variations: i) For the $W$=15nm UTB film there is strong anisotropy between the [110] direction (arrow, $x$-axis) and the [001] ($y$-axis). ii) When the thickness reduces (Fig. 2d), the contour in the [110] direction changes, acquiring a larger curvature (the contours now point towards the center). iii) The change in the energy contours is anisotropic, i.e. no significant changes are observed to the contours along the [001] direction (*y*-axis). This behavior is a consequence of the anisotropic shape of the heavy-hole subband. Detailed explanations on the effect of confinement on the band shapes are provided in Refs [18, 19, 25] and we refer the reader to those works.

In order to extract the UTB channel velocities, we plot the velocity versus carrier concentration at high $V_D$, (as shown in Fig. 3a for the (100)/[110] UTB channels), for layer thicknesses from *W*=3nm (black-solid) to *W*=15nm (black-dot). We then pick the low carrier concentration, most left value for each case. This represents the non-degenerate case, where Boltzmann statistics apply. Figure 3b shows the UTB hole velocities for the [100], [110] and [111] transport and different surface confinement orientations versus the UTB film thickness. Strong anisotropic behavior is observed in both, surface, and transport orientations. The (110) and (112) surfaces provide the highest velocities, whereas the (100) surfaces the lowest. Overall, the (110)/[110] UTB channels have the highest velocities, whereas the (100)/[110] the lowest ones. The fact that they share the same transport orientation but different confinement, indicates that the effect of the surface confinement is more prominent than the effect of transport orientation.

It is also evident that as the thickness of the UTB scales down, the carrier velocities increase. Figure 3c shows the relative change in the hole velocities of the different UTB categories, normalized to their highest value (at the smallest thickness). The largest variations are observed for the (110) and (112) surfaces. Especially in the case of the (110)/[110] and (112)/[111] UTB channels, the velocity increase can reach up to ~35%. The (100) surface UTB channels still benefit from thickness scaling, but only by a small amount ~10%, and only at thicknesses below *W*=5nm. In reality, however, surface roughness scattering is stronger for film thicknesses below 5nm and benefits for this case might, or might not be observed.

The dispersions that determine the velocities in these channels are mixtures of subbands originating from heavy-hole, light-hole and split-off bands, and in addition include band mixing, valley splitting, and quantization features. These cannot be separated in a trivial manner. A single value of an "approximate" transport effective mass that will account for the combination all various bands can be a useful quantity that would partly describe the electronic properties of UTB channels. It can potentially provide a reasonable metric when it comes to comparison with different channel materials. An "approximate" transport effective mass can be extracted using the carrier velocities of the UTB channels in the non-degenerate limit as $m_t^* = 2k_BT/(\pi v^2)$ [11, 32]. This is plotted in Fig. 3d for all the UTB cases considered. The trend is the inverse of what the velocities follow in Fig. 3b. The (110)/[110] channel has the lowest "approximate" hole mass, and under strong scaling at $W$=3nm, it reduces to $m^* \sim 0.19m_0$, the same as the value of the electron transverse mass in the conduction band. This is an indication that such a channel will exhibit a higher phonon-limited mobility compared to bulk p-type devices (since $\mu = q\tau/m^*$). Indeed, for [110] NWs of diameter $D$=3nm, mobility calculations suggest very high phonon-limited mobilities [33, 34], similar to the bulk n-type mobility and certainly higher than the bulk p-type mobility. Of course, the effect of surface roughness scattering (SRS) will be stronger in UTB channels and degrade the mobility. This reduction in the "approximate" effective mass, however, can be a mechanism to partly compensate for the effect of SRS.

The hole velocity behavior of the UTB channels originates from the underlying dispersions. In Fig. 4, the envelope dispersions (highest subbands) for two surface orientations, the (110) and (100), are plotted. Figure 4a shows the first subband (envelope) of the (110)/[110] UTB channels versus $k_x$ with $k_y$=0 (where $x$ is the transport and $y$ is the transverse direction), as the width reduces from $W$=15nm to $W$=3nm (in the direction of the arrow). Equivalently, these are bands along the arrows of Fig. 2c and 2d. As shown in Fig. 4a, scaling of the (110) surface results in bands with larger curvature, which provide increased carrier velocities (Fig. 3b). The same is observed for the (110)/[100] channels of Fig. 4b, but at a smaller degree (equivalently bands along the $y$-axis [100] in Fig. 2c and Fig. 2d). The (110) confinement provides light subbands due to the anisotropic behavior of the heavy-hole band [18, 19]. The [110] transport direction takes full advantage of that, whereas the [100] direction less. We mention that increased carrier velocities are observed in any case where structure quantization picks bands from high curvature regions of the bulk Brillouin zone. This can also be observed at a smaller degree in the [110] orientation for n-type NWs under cross section scaling [25], as well as in other materials where the valence or conduction bands have strong anisotropic shapes, i.e III-V materials such as InAs or InSb.

Figures 4c and 4d show the envelope subbands versus $k_x$ (again for $k_y$=0) for UTB films with (100) surface confinement, in the [110] and [100] transport orientations, respectively. Equivalently, these are bands along the diagonal lines at 45° in Fig. 2a-b, and along the arrows of Fig. 2a-b, respectively. No variation is observed in the envelopes of the bands as the width reduces, which justifies the very small velocity variations in Fig. 3c for these channels. The slight increase in the velocities for the $W$=3nm UTB channels can be explained by looking at the highest four subbands. The insets of Fig. 4c and 4d show the highest four subbands of the $W$=15nm (red) and $W$=3nm (black) channels. In the wider channels all four bands have a similar curvature. The first and second bands of the thinner channels also have similar curvature to the wider channels. The third and fourth bands of the thinner channels, on the other hand, have a larger curvature. These two bands provide a slightly higher hole velocity as the width of the UTB reduces.

## IV. Discussion and Design Considerations

The strong anisotropic behavior of the p-type UTB channels points toward design optimization strategies. In the case of high performance quasi-ballistic MOSFET devices, high hole velocities will improve the performance in terms of on-current ($I_{ON}$). In the case of long channel diffusive transport devices, high carrier velocities can improve the channel mobility. Our results indicate that for p-type devices, the (110)/[110] channels will have the highest velocities, followed by the (112)/[111], and then by the (110)/[100] channels. On the other hand, the (100)/[110] and (100)/[100] channels lack on performance. Optimal choices of surface and transport orientation, as well as thickness scaling can, therefore, improve carrier velocities. The calculated velocity trends with respect to the channel orientation are in good agreement with experimental measurements of hole mobility in MOSFET devices of various surface and transport orientations [12, 14, 28]. Even the relative differences between the various cases are within reasonable agreement. A comparison between simulation and experiment indicates that the carrier velocity could be correlated to the low field mobility as also indicated in Ref. [11] for n-type UTB channels. Other simulation studies on mobility of some cases of p-type UTB channels in various orientations [13] also agree with the trends calculated here. We mention here that a common practice to improve the performance of p-type (100) MOSFET devices is the introduction of strain [35]. Whether these results will hold in the presence of strain is something still to be examined, but what we describe here can serve as an additional performance optimization mechanism.

In reality, however, UTB layers with thickness below 6nm would suffer from enhanced surface roughness scattering (SRS) [36]. Channels that provide improved carrier velocities as the channel width reduces could partially compensate for SRS and still provide attractive electronic properties. This is the case for (110)/[110] and (112)/[111] channels, in which the hole velocity increases by ~35% as the channel width reduces down to $W$=3nm. For example, an effective way to design high efficiency nanostructured thermoelectric devices is to scale the feature sizes in order to increase phonon-boundary scattering and reduce the lattice thermal conductivity. High electronic conductivity, though, is still needed. The (110)/[110] or (112)/[111] p-type channels might be more optimal for such applications compared to channels with other surface/transport orientations.

For high performance, quasi-ballistic transistor applications, on the other hand, what is needed is high on-current $I_{ON}$. The $I_{ON}$ in ballistic devices is given by the product of charge times velocity $I_{ON} = qn \times v_{inj}$. The charge in nanoscale devices is still mostly controlled by the gate electrostatics and the oxide capacitance. With the same oxide capacitance (oxide thickness and dielectric constant), all 2D UTB channels considered will have similar charge density at the same inversion conditions, irrespective of orientation and channel thickness. Some differences in the charge can arise from differences in the quantum capacitance of the channels. The low quantum capacitance can reduce the total gate capacitance by a factor of ~30%. Still, however, this reduction is very similar for all the channels we are considering for reasons explained in Refs [18, 25] and do not produce significant variations in the charge density between the different channels. Figure 5a shows the variation of the charge versus gate bias for (100)/[110] UTB channels of different thicknesses. Indeed, the change in the channel does not vary significantly, irrespective of the thickness. Figure 5b shows the charge density in the UTB channels of all orientations considered as a function of their thickness at high inversion conditions $V_G$=1V and $V_D$=0.5V. Only small charge variations between orientations and thicknesses are again observed. The inset of Fig. 5b shows the charge values normalized to their highest value (that of the wider channels). The maximum charge reduction with thickness scaling is only ~13%. This charge variation behavior is different from what is observed for NWs, for which the oxide capacitance, and correspondingly the charge, decreases linearly with diameter scaling [15].

Figure 5c shows the ballistic on-current for all orientations and channel thicknesses. The current follows the velocity trend of Fig. 3b. The normalized currents to their highest value shown in Fig. 5d also indicate in all cases very similar current variation trends as the velocity variation trends shown in Fig. 3c. The increase in the current for the (110)/[110] and (112)/[111] channels as the thickness scales reaches up to ~35%. The velocity trends, therefore, directly relate to ballistic $I_{ON}$. For high performance, close-to-ballistic MOSFET devices, the (110) or (112) surfaces provide channels with higher current densities. If one, however, is interested more in reducing thickness related device-to-device performance fluctuations in the expense of reduced performance, the (100) surfaces are the ones more tolerant to thickness fluctuations. This behavior is different from what we have earlier reported for NWs in Ref. [15]. In NW channels, the capacitance and charge vary linearly with diameter. Therefore, as the diameter reduces, the $I_{ON}$ also reduces (but less so for the NW cases where $v_{inj}$ increases).

We mention here that in this work we have not considered the effect of potential variations in the cross section of the UTB channel. Including the electrostatic potential can have some effect on the magnitude of the total gate capacitance as well as the on-currents [37]. However, we do not expect this to largely impact the trends we present for the on-current variations.

Finally, for the completeness of this analysis, we mention that although the (110) surface is beneficial for holes and the (100) surface is not, in the case of n-type UTB devices this behavior is reversed. As shown in Fig. 6a, for electrons the (100)/[110] and (100)/[100] channels are more beneficial than the (110)/[100] channels, and especially the (110)/[110] ones that indicate the worst performance in terms of carrier velocities. These results also follow the trend of experimental mobility measurements in Ref. [12, 28]. Although in the multi-valley transport n-type case it is less trivial to make a direct connection of the mobility to the average carrier velocity, a correlation is evident as it has also been discussed elsewhere [11]. The (100) surfaces utilize the light transverse masses ($m^*=0.19m_0$) of silicon conduction band ellipsoids more, whereas the (110) surfaces utilize the heavier masses of the rotated ellipsoids more ($m^*=0.55m_0$) [17, 25]. The n-type (100) surface channels also have the largest velocity increase as the UTB channel thickness reduces (Fig. 6b). The reason for this increase is that (100) quantization causes a stronger upward energy shift of the heavier transport (but lighter quantization) mass off-$\Gamma$ valleys, compared to the projected $\Gamma$ valleys of the Si UTB channel. On the other hand, the electron velocities of n-type (110) surfaces increase only slightly (in [100] transport), or even decrease (in [110] transport) as the thickness reduces (Fig. 6b). This depends on whether the light transport mass valleys are shifted higher in energy than the heavier ones. A slight variation in the subband curvature is also observed with quantization [15, 25], but the electron velocities are mostly controlled by the relative placement of the heavy/light mass valleys, rather than curvature variation as in the case of hole velocities. Nevertheless, the maximum variation in $v_{inj}$ is less than ~20%, a factor of ~2X weaker than in p-type channels. Hybrid orientation technologies utilize both n-type and p-type channels on the same substrate [12, 14]. In such cases, careful design considerations will be needed to ensure high performance for both, holes and electrons.

We believe that this study can provide useful guidance in choosing p-type UTB film thickness, transport, and confining surface orientations for performance improvement.

## V. Conclusion

The hole velocity in p-type silicon UTB channels is calculated using atomistic tight-binding considerations and semiclassical ballistic transport. The results present a comprehensive analysis of hole velocities in various transport and surface orientations for channel thicknesses varying from $W$=15nm down to $W$=3nm. The hole velocity is strongly anisotropic depending on the channel confinement and transport orientations. The (110) and (112) confinement surfaces offer the highest velocity performance, whereas the (100) surface the lowest, with ~30% lower hole velocities. In addition, (110) surface scaling further increases hole velocities by up to ~35%. Transport orientation can also be important, with the (110)/[110] channels being the optimal choices for high hole velocities and ballistic on-currents (whereas the (100)/[110] channels are the best for electron velocities). The hole velocity behavior originates from the large curvature variations in the dispersions with both, orientation and confinement. This is a consequence of the anisotropic nature of the silicon bulk heavy-hole band. The velocity trends translate directly to ballistic on-current trends. They also agree with recent experimental mobility measurements for p-type UTB films and MOSFET devices in various surface and transport orientations. Our analysis connects these experimental observations directly to bandstructure features. Thus, it can provide insight and guidance towards optimization of p-type UTB devices for a variety of electronic transport applications.

## Acknowledgements

This work was supported by the Austrian Climate and Energy Fund, contract No. 825467. The simulations in this work can be duplicated with Bandstructure Lab on nanoHUB.org [38].

Figure 1:

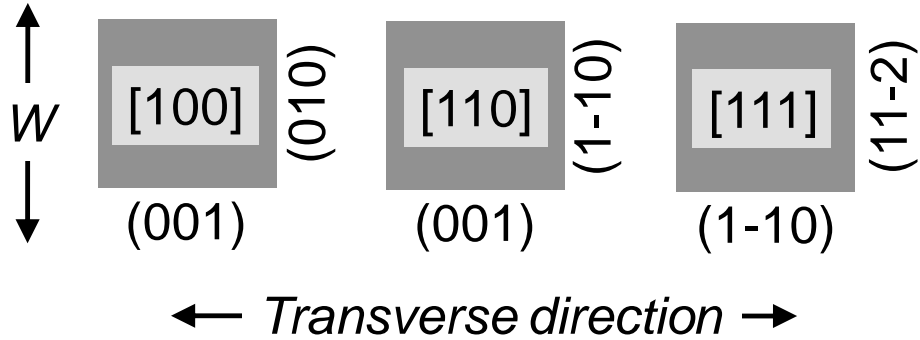

Figure 1 caption:

The UTB film orientations considered in this work are indicated on the squares that represent the channels. The transport orientations are noted in the center of the square and the different surfaces are denoted on the sides. *W* is the confinement width of the channel.

Figure 2:

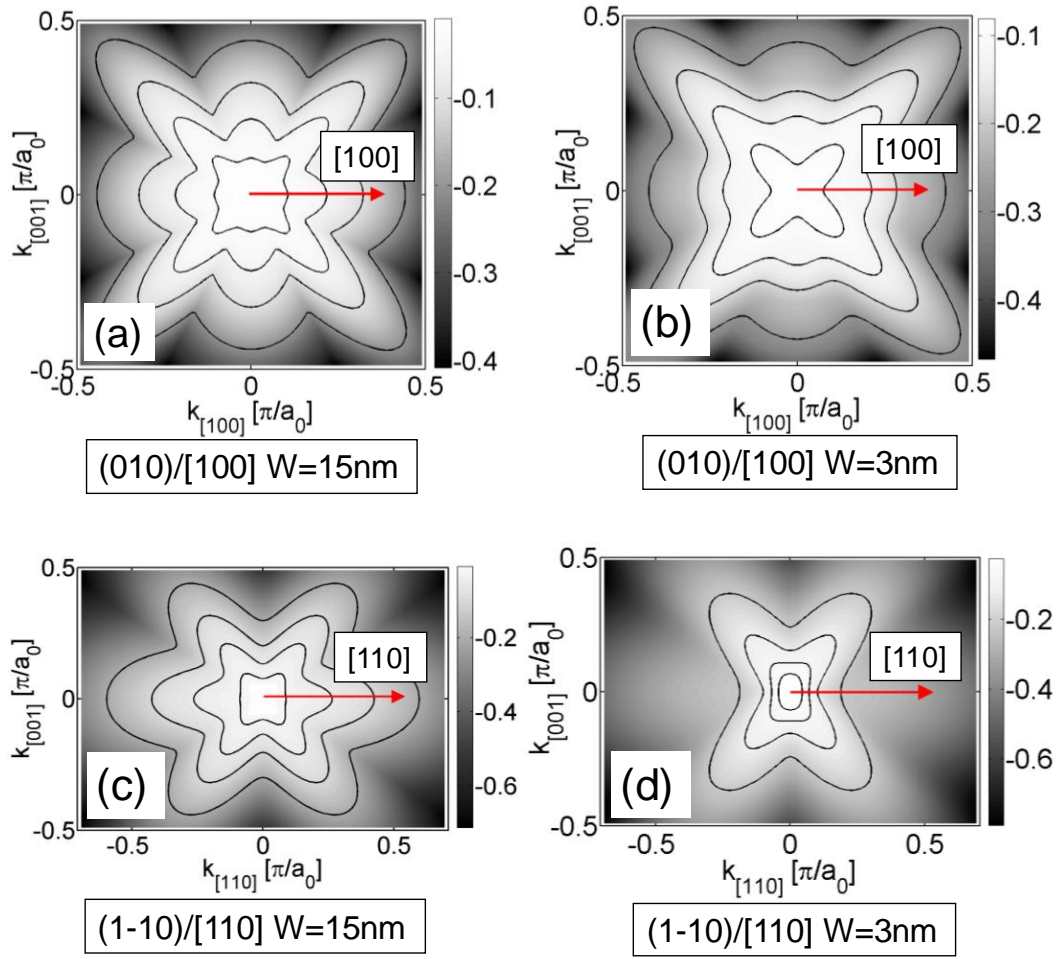

## Figure 2 caption:

Energy surfaces of UTB layers for different thicknesses (*W*) and orientations: (a) (010) surface orientation with *W*=15nm. (b) (010) surface orientation with *W*=3nm. (c) (1-10) surface orientation with *W*=15nm. (d) (1-10) surface orientation with *W*=3nm. The contour lines (from the center outwards) represent energy contours at 0.02eV, 0.05eV, 0.1eV and 0.2eV below the top of the valence band.

Figure 3:

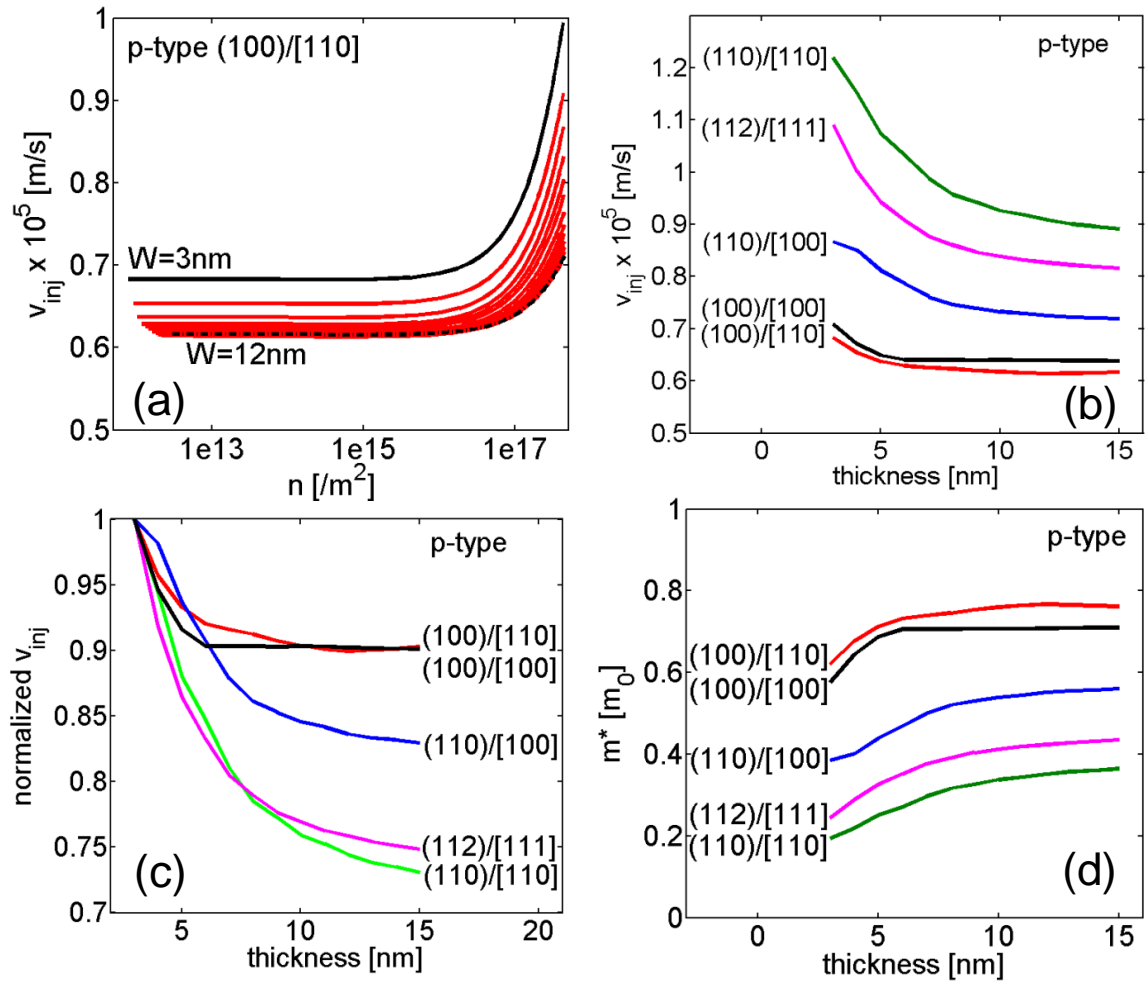

Figure 3 caption:

Hole velocities in UTB channels of different surface and transport orientations versus the layer thickness. (a) p-type (100)/[110] channel velocities versus the carrier concentration for different layer thicknesses. (b) The hole velocities of the UTB channels. (c) The hole velocities normalized to their largest value. (d) An estimate of an "approximate" transport effective mass based on the velocity values and non-degenerate limit considerations.

Figure 4:

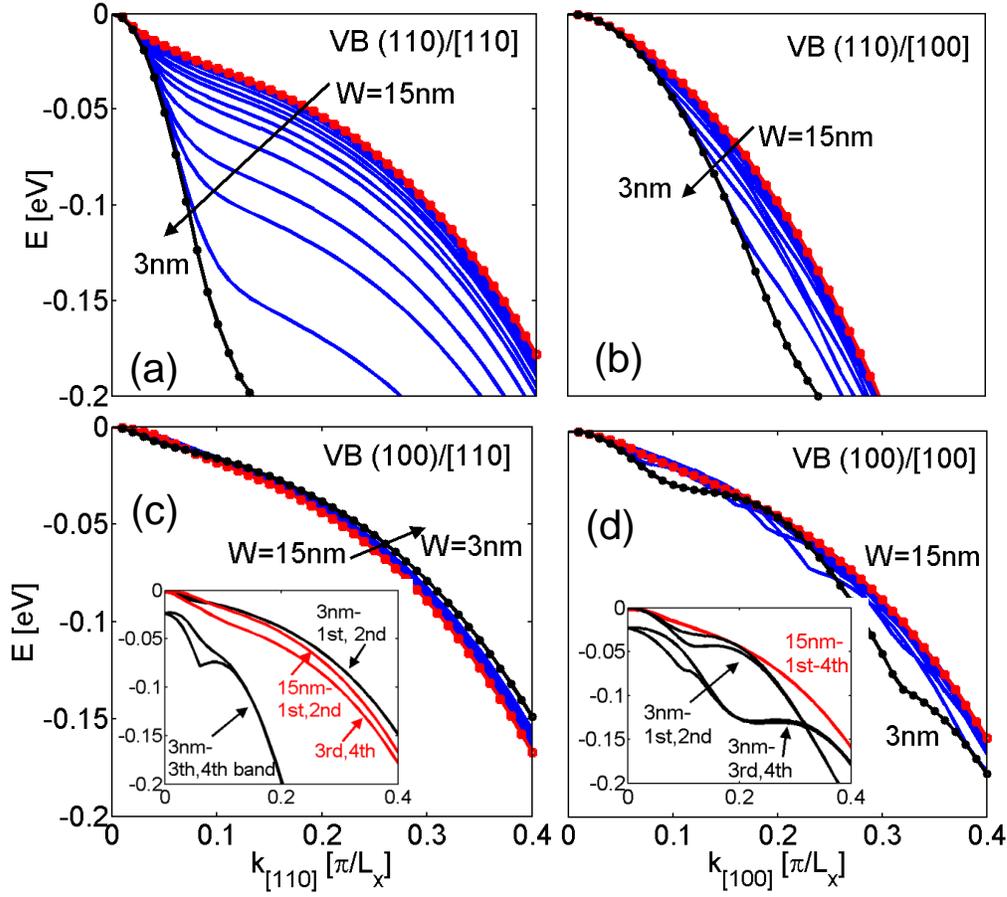

## Figure 4 caption:

The first subband (envelope) of different UTB channels for thicknesses from $W$=15nm down to $W$=3nm versus $k_x$ with the transverse $k_y$=0. $L_x$ is the length of the unit cell in the transport direction, $L_x = a_0/\sqrt{2}$ for (a-b), and $L_x = a_0$ for (c-d), where $a_0\sqrt{3}/4$ is the silicon bond length. (a) (110)/[110] channels. (b) (110)/[100] channels. (c) (100)/[110] channels. (d) (100)/[100] channels. The arrows point toward the direction of thickness reduction. Insets of (c) and (d): The first four subbands (envelopes) for the $W$=3nm (black) and $W$=15nm (red) channels.

Figure 5:

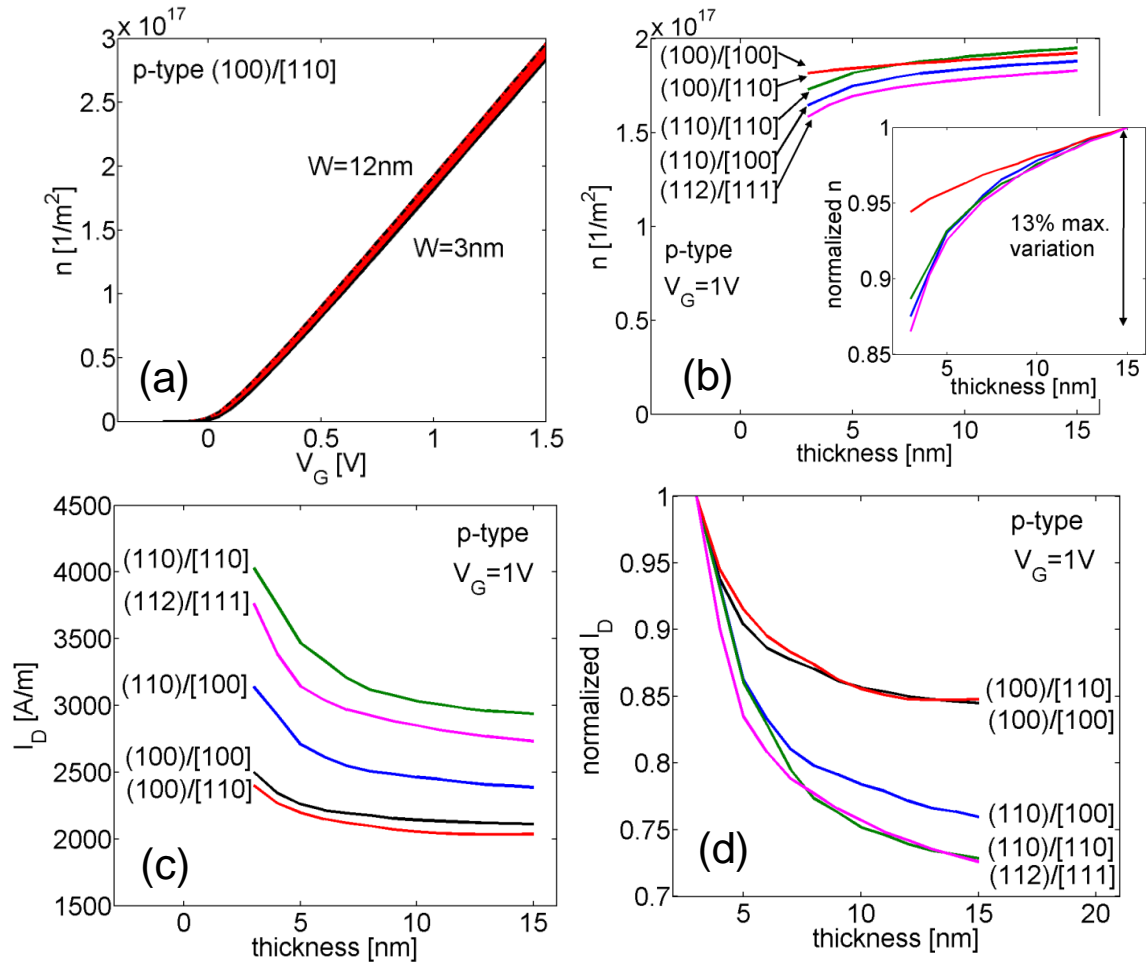

Figure 5 caption:

Hole charge density and ballistic currents for UTB channels of different surface and transport orientations versus the layer thickness. (a) p-type (100)/[110] channel charge versus gate bias for different layer thicknesses. (b) The charge in the UTB channels at inversion $V_G$=1V, $V_D$ = 0.5V. Inset: The charge values normalized to their highest value. (c) The ballistic on-currents for the UTB channels for $V_G$=1V, $V_D$=0.5V. (d) The ballistic on-currents normalized to their highest value.

Figure 6:

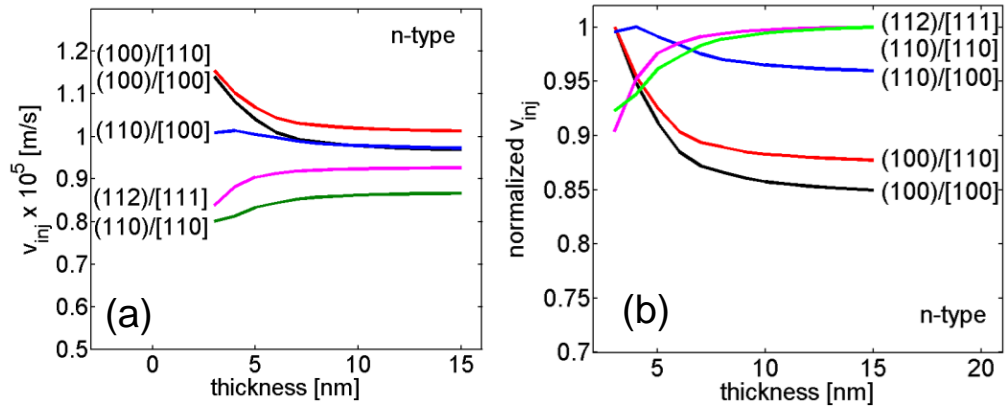

Figure 6 caption:

Electron velocities in n-type UTB channels of different surface and transport orientations versus layer thickness. (a) The electron velocities of the UTB channels. (b) The electron velocities normalized to their highest value.